# Symbolic sensors : one solution to the numerical-symbolic interface

E. Benoit  L. Foulloy

Laboratoire d'Automatique et de MicroInformatique Industrielle, Université de Savoie
41 Avenue de la plaine, BP 806, 74016 Annecy Cedex, France

**According to the evolution of control systems, the nature of information become more complex. Artificial intelligent technics use and produce symbolic information, therefore there is a need for sensors which deliver this kind of information.**

**This paper introduces the concept of symbolic sensor as an extension of the smart sensor one. Then, the links between the physical world and the symbolic one are introduced. The creation of symbols is proposed within the frame of the pretopology theory. In order to adapt the sensor to the measurement context, a learning process has been used to provide an adaptive interpretation of the measurement. Finally, an example is presented in the case of a temperature measurement.**

# Introduction

In every control systems, a part of the signal processing is attached to the sensors. Therefore, as controllers become more complex, a part of the signal processing is transferred to the sensor. Analog sensors integrate it in the conditioner, while smart sensors have computation devices to process the signal. The most important difference between this two king of sensors is the nature of informations exchanged with the controller, it fixes the maximum decentralization of the control system.

Nowadays, it is generally admitted that the characteristic functionality of a smart sensor should be its ability to communicate with a communication bus or network, to verify the correctness of the measurement and to adapt itself when the environment is changing (Burd [1], Giachino [2], Bois [3]). A possible description of a smart sensor is given by Favenec [4] (see Fig. 1.). We consider that an intelligent sensor should own four mechanisms: perception (the main part of the sensor), communication, learning and faculty of reasoning.

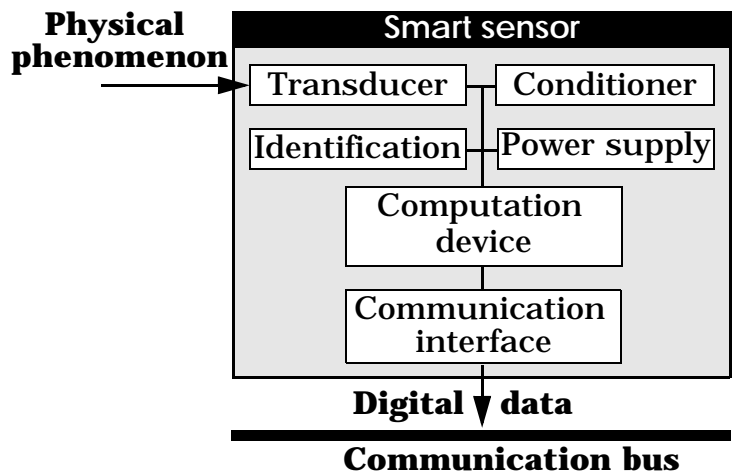

Fig. 1. Internal scheme of a smart sensor.

Applications of artificial intelligent technics like expert systems and qualitative control, use symbolic informations. Then, the decentralization of a part of the decision process induces adapted perception organs. For example the Order of Magnitude model uses symbols like "negative medium" or "positive small" [5], therefore there is a need for sensors that return this kind of symbols.

**We define the symbolic sensor as a smart sensor which is able to work and to provide symbolic informations relative to the measurement.**

This new property allows the sensor to make decisions about the measurement and the context, it adds the possibility of reasoning that is commonly used by human beings in usual intelligent activities.

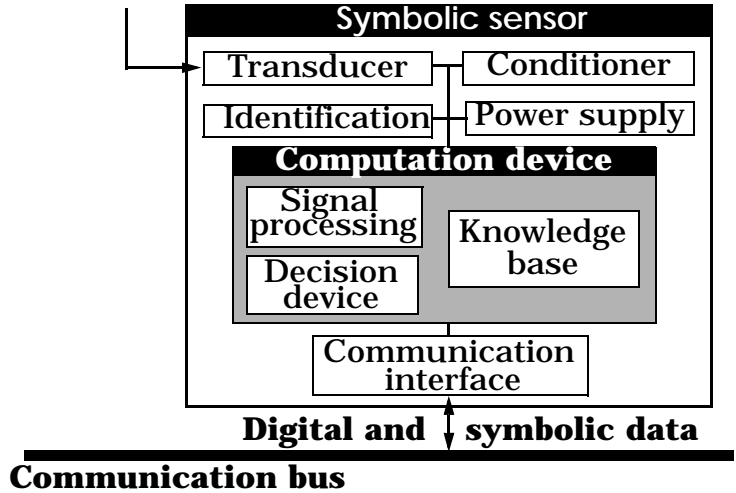

Fig. 2. Internal scheme of a symbolic sensor.

## Translation, concept and interpretation

In order to realize a numeric to symbolic link, we define a numerical domain $E$ as the set of the measurement and a symbolic one $L$ as a set of symbols characterizing the measure. The meaning of a symbolic value will be called **a translation** and be defined as an injective application from the symbolic set to the set of the subsets of the numerical domain (injectivity insures that two identical symbols have the same translation). A generalization of the concept of translation was developed by Luzeaux [6].

$$\tau: L \to P(E)$$

The association of a symbolic value and its translation is called **a concept**. The symbolic measurement will be obtained by means of a new relationship from the numerical domain to the symbolic one, called **an interpretation**.

$$\iota: E \to L$$

The relationship between the translation and the interpretation is summarized in Fig. 3.

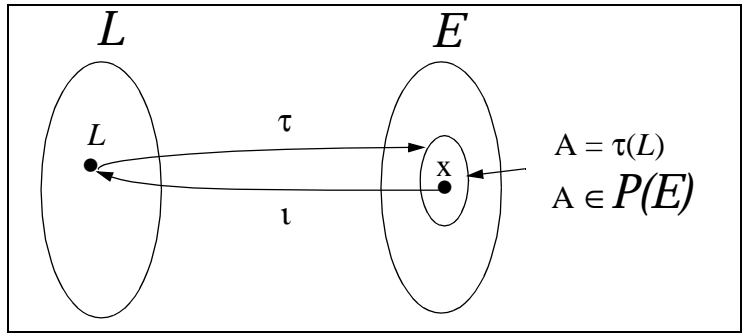

Fig. 3. Relationship between a translation and an interpretation

The definition of the concepts for any symbol induces the definition of the translation. The translation can be unconstrained, then, for any symbol $L$, $\tau(L)$ can be any subset on $E$. In the following of this paper, we study a kind of symbolic sensor which impose several conditions on the translation:

• The intersection between the translations of two symbols must be empty.

• The numerical domain is an ordered set with an order relation noted $<_E$.

• The translation of any symbol is an interval:
$\forall L \in \mathcal{L}$, $\forall x, y \in \tau(L)$ such that $x <_E y$, $\forall z \in E$
$x <_E z <_E y \Rightarrow z \in \tau(L)$

Then, the interpretation can be define as follow.
$$L = \iota(x) \text{ if } x \in \tau(L)$$

This relationship is an application when $\tau(L)$ is a partition on $E$. This assumption will be used in the following.

## The pretopology formalism

### a) Introduction

We give here several notations used in the paper: $A \cup B$ (resp. $A \cap B$) will denote the union (resp. the intersection) of two sets A and B. The difference of sets A,B will be represented by A- B. The complementation of a set A will be denoted by c(A). The positive integer card(A) denotes the cardinal number of the set A. If f is a relation on E, A a subset of E and n an integer up to 0 then

$$f^n(A) = f(f^{n-1}(A)) \qquad f^1(A) = f(A)$$

Originally used for morphological image analysis, the pretopology formalism gives operators such as dilatation and erosion working on sets. Before using this mathematical tool on concepts, let us recall the principles of the pretopology described by Emptoz [7]:

Let E be a set, for any x in E, is given a family $B(x)$ of subsets of E which all contain x. This family is called a structuring base. Given $B(x)$ for any x in E, we say that E is embedded with a pretopological structure or simply a pretopology. Two basic operators, the adherence and the interior are defined as follows.

Let A be a subset of E
$adh_B(A) = \{x \in E \ / \ \forall B, B \in B(x), B \cap A \neq \emptyset\}$
$int_B(A) = c(adh_B(c(A)))$

Example: $E = \mathbf{Z}^2$, $x = (i, j)$, $B(x) = \{B1, B2\}$ ($\mathbf{Z}$ is the set of signed integers) with:
B1 = {(i-1,j),(i,j),(i+1,j)}
B2 = {(i, j-1),(i, j), (i, j+1)}

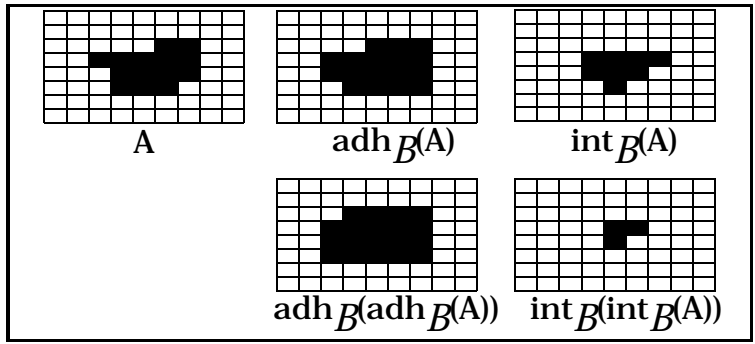

A     $adh_B(A)$     $int_B(A)$

$adh_B(adh_B(A))$   $int_B(int_B(A))$

Fig. 4. Example using interior and adherence.

Assuming that pred(x) is the element which precedes x and succ(x) the element which follows x in a totally ordered discrete set, the two following structuring bases are defined:

$Bsup(x) = \{\{pred(x), x\}\}$
and
$Binf(x) = \{\{x, succ(x)\}\}$

Let us give several examples of the effect, in the set E={0,...,8}, of the adherence and interior operators using $Bsup(x)$ and $Binf(x)$.

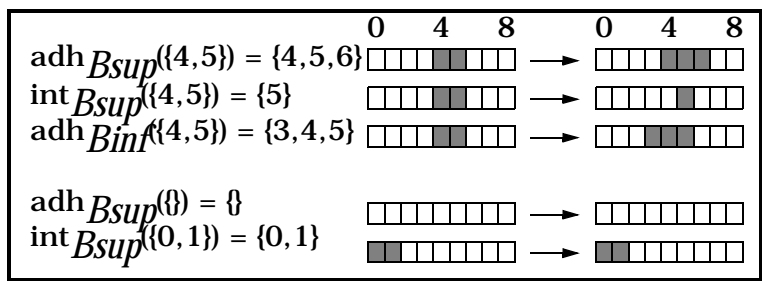

$adh_{Bsup}(\{4,5\}) = \{4,5,6\}$
$int_{Bsup}(\{4,5\}) = \{5\}$
$adh_{Binf}(\{4,5\}) = \{3,4,5\}$

$adh_{Bsup}(\{\}) = \{\}$
$int_{Bsup}(\{0,1\}) = \{0,1\}$

Fig. 5. Action of int and adh with $Bsup$ and $Binf$

These two bases will be used in the following.

## b) The generic concept and the creation of new concepts

The generic concept corresponds to a part of the numerical domain which is important for the global task that is to be realized with the sensor. Let us give an example. Assume that a thermometer is used for a swimming-pool. It makes sense to define a generic concept **temperature is correct** around 23° C. Now, if the sensor is used into a deep freeze, the generic concept **temperature is correct** should be around -18° C. According to our approach, the translation associated to the generic concept is an interval on the numerical domain. Let us define the translation associated to generic concept **temperature is correct** for the swimming pool example by τ(temperature_is_correct) = {21, 22, 23, 24, 25}.

Obviously, it is a tedious task to specify each translation of the symbols. Let us introduce the operators that are used to generate new concepts. Our problem is to find operators that act in the symbolic domain and their definition in the numerical domain in order to compute the new concepts. First, let us define a mapping function $F$ in the symbolic domain $L$.

$F: L \to L$
$L_1 \mapsto L_2 = F(L_1)$

We can now define the associated mapping function in the numeric domain $E$.

$f_F: P(E) \to P(E)$
$\tau(L_1) \mapsto \tau(L_2) = \tau(F(L_1)) = f_F(\tau(L_1))$

The relationship between functions $F$ and $f_F$ is given in Fig. 6.

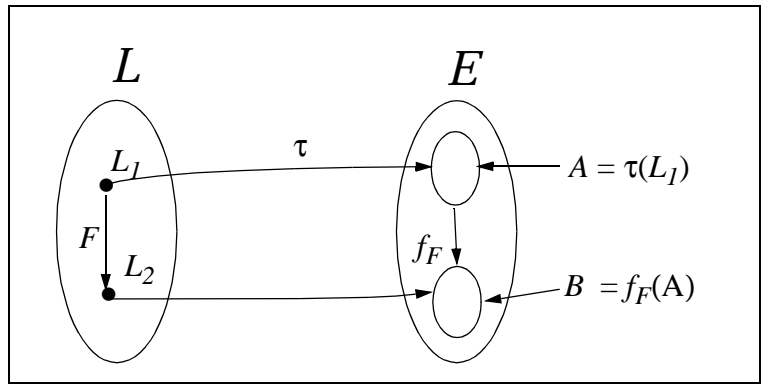

Fig. 6. Relationship between $F$ and $f_F$.

Assuming that $A \in P(E)$, $L_1$ and $L_2 \in L$ such that $A = \tau(L_1)$, four mapping functions have been defined with the help of the pretopology formalism.

$$L_2 = \textbf{more}(L_1) \quad f_{more}(A) = adh^{card(A)}_{Bsup}(A) - A$$

$$L_2 = \textbf{less}(L_1) \quad f_{less}(A) = adh^{card(A)}_{Binf}(A) - A$$

$$L_2 = \textbf{below}(L_1) \quad f_{below}(A) = \lim_{n \to \infty} adh^{n}_{Binf}(A) - A$$

$$L_2 = \textbf{above}(L_1) \quad f_{above}(A) = \lim_{n \to \infty} adh^{n}_{Bsup}(A) - A$$

Let us give an example for a temperature sensor on a discrete measurement set (temperature is given in °C) $E = \{0,1,...,38,39\}$. First of all, we define the generic concept **good** such as $\tau(\textbf{good}) = \{18, 19, 20, 21, 22\}$. Then, we inform the sensor about the semantics between the generic concept and new concepts.

**cold** = **below(good)** and **warm**=**more(good)**.

The sensor uses this information to build its concepts (see Fig. 7.).

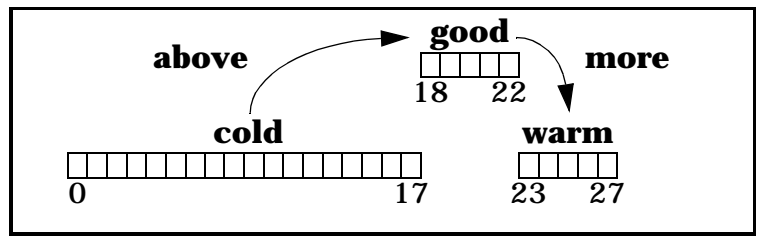

Fig. 7. Concepts building.

### c) Adaptation of the interpretation

In the previous sections, it has been shown how to build automatically with several operators one relationship between the numerical and the symbolic domains. It seems obvious that the definition of translations is the difficult point. A symbolic sensor that is concerned with a temperature measurement problem can interpret the measure "temperature=15°C" as a cool temperature. According to the context, the interpretation can be false: joined with a swimming-pool, the temperature is cold. Joined with a refrigerator, it is very hot. Furthermore, the interpretation can differ with the interlocutor. This last point is very important during the knowledge acquisition time when one wants to implement an expert system. Therefore, there is a need to configure the sensor according to the context (envi-ronment, measurement job) and the interlocutor. The ***cooperative configuration*** is based on a learning method with a teacher (control system, human or aggregation sensors) who interprets correctly a measure in a defined context.

The principle of the cooperative configuration is given in Fig. 8. It is based on the qualitative comparison between the interpretation of the sensor (**sc**) and the interpretation of the professor (**sp**) for the same numerical input **x**. The qualitative comparison **e** takes its values in the set of signs **S** = {-, 0, +} usual in qualitative analysis. The result is obtained from the order relation on $L$.

$e = +$ *if* $sp >_L sc$
$e = -$ *if* $sp <_L sc$
$e = 0$ *if* $sp =_L sc$

The qualitative comparison is used to synthesize a symbolic input **u** for the sensor which takes its value in the set **M** = {**increase**, **decrease**, **maintain**}. This symbolic input is used to change the translation of each symbolic value and, of course, the sensor interpretations. The idea is to change the translation in order to obtain a qualitative feedback of the sensor interpretation according to the professor interpretation.

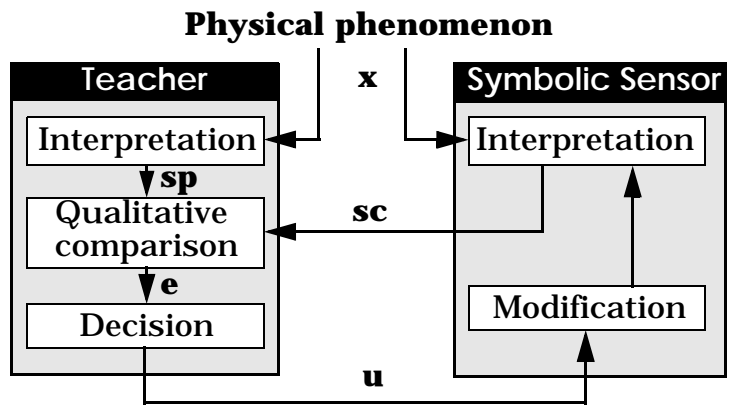

Fig. 8. principle of cooperative configuration.

The qualitative feedback can be easily obtained with the following control strategy.

*if* $e = +$ *then* $u =$ **increase**
*if* $e = -$ *then* $u =$ **decrease**
*if* $e = 0$ *then* $u =$ **maintain**

The semantics of the control actions on **u** are respectively to increase, decrease or maintain the sensor interpretation. Let $\iota^k(x)$ be the inter-pretation of the measurement *x* at the learning cycle *k* and $\tau^k(L_i)$ the translation of the lexical value $L_i$ at the learning cycle *k*, the control actions are such that:

$$\iota^{k+1}(x) = increase(\iota^k(x)) \geq_L \iota^k(x)$$
$$\iota^{k+1}(x) = decrease(\iota^k(x)) \leq_L \iota^k(x)$$
$$\iota^{k+1}(x) = maintain(\iota^k(x)) =_L \iota^k(x)$$

Now, in order to complete the approach, the function increase, decrease and maintain have to be defined. The function maintain is obvious since no action is performed. Let $L \in \mathcal{L}$ be the interpretation of the measurement *x*, $L = \iota^k(x)$, then the functions *increase* and *decrease* are defined as follows from the interior operation (another definition from the adherence operator can easily be obtained).

$\iota^{k+1}(x) = increase(L)$:
$$\tau^{k+1}(L) = int_{Binf}(\tau^k(L))$$
$$\tau^{k+1}(succ_L(L)) = \tau^k(succ_L(L)) \cup (\tau^k(L) - \tau^{k+1}(L))$$

$\iota^{k+1}(x) = decrease(L)$:
$$\tau^{k+1}(L) = int_{Bsup}(\tau^k(L))$$
$$\tau^{k+1}(pred_L(L)) = \tau^k(pred_L(L)) \cup (\tau^k(L) - \tau^{k+1}(L))$$

The application of such functions on a lexical set with five symbolic values at the learning cycle *k* is provided in Fig. 9. to Fig. 11.

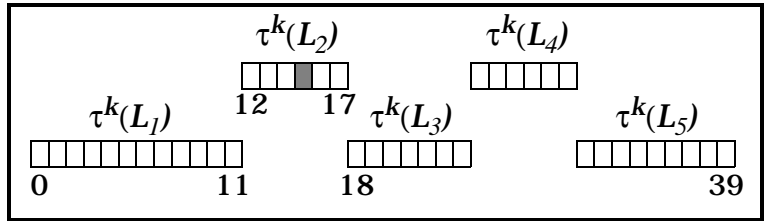

Fig. 9. Translation at the learning cycle *k*.

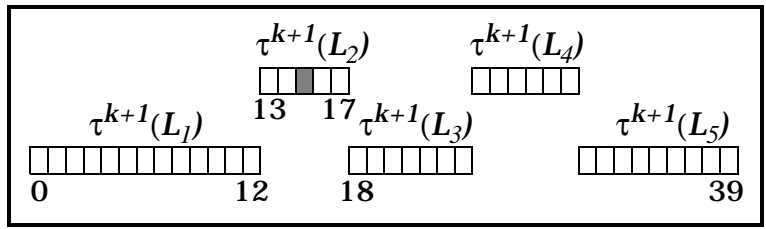

Fig. 10. New translation after the application of decrease for x=15.

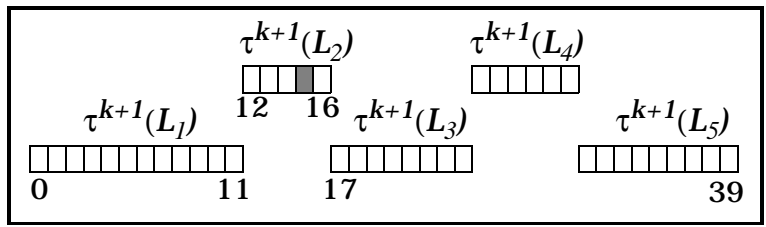

Fig. 11. New translation after the application of increase for x=15.

## Application

We illustrate This formalism with an example of cooperative configuration of a temperature sensor. At first, we build a symbolic sensor which have the symbolic domain $L$={ **cold**, **cool**, **good**, **warm**, **hot**} and a numerical domain $E$=[$0^oC$, $39^oC$] the discrete set of integer value, in Celsius, of temperature between $0^oC$ and $39^oC$:

We give the following informations to the sensor:

- the generic concept is **good** associated with **[18,22]**.
- **warm** is **more** than **good**.
- **hot** is **above warm**.
- **cool** is **less** than **good**.
- **cold** is **below cool**.

The Fig. 12. shows the concepts built by the sensor.

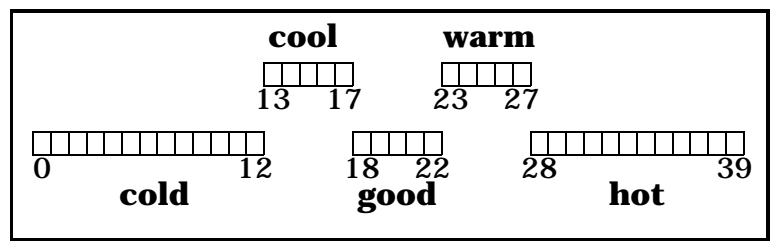

Fig. 12. The five concepts.

During a learning cycle, the sensor makes a measurement and give it to the teacher (the author in this case), this one makes a qualitative comparison with its interpretation and gives the appropriate action to the sensor. The modification of the perception of the environment is performed around four points in the measurement set ($28^o$, $25^o$, $17^o$ and $12^o$) during 10 learning cycles. The five first learning cycles are detailed:

Learning cycle 1
   the temperature is              : 25ºC

the sensor interpretation is : warm
the teacher action is : decrease

Learning cycle 2
the temperature is : 25°C
the sensor interpretation is : warm
the teacher action is : decrease

Learning cycle 3
the temperature is : 25°C
the sensor interpretation is : warm
the teacher action is : decrease

Learning cycle 4
the temperature is : 25°C
the sensor interpretation is : good
the teacher action is : maintain

Learning cycle 5
the temperature is : 28°C
the sensor interpretation is : hot
the teacher action is : decrease

The Fig. 13. shows the concepts after the configuration.

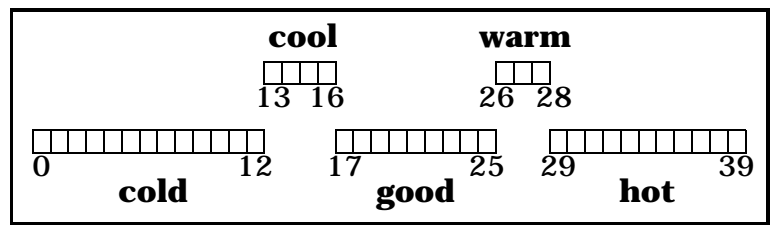

Fig. 13. Result of the configuration.

## Conclusion

After the integration of the analog to digital conversion and the signal processing into intelligent sensors, the integration of a part of the decision process and of the knowledge into the symbolic sensor represents a logical evolution of the decentralization effort of the supervisor system tasks. The symbolic sensor has a perceptive knowledge, materialized through the segmentation of the measurement domain, and a semantics knowledge (the mean of words like "more than" or "increase").

The cooperative configuration enables the acquisition of this knowledge without knowing the relation between the measure, the context and the interpretation. So, the possibility to have a symbolic sensor which can interpret human concepts like comfort or danger can be considered. The expert system, integrated in the sensor, is able to manipulate the knowledge base and the perceptive knowledge, it can take a decision about the sensor itself or about the environment. For example, it can see a failure on itself or determine the context by a communication with the other sensors and the system.

This paper has presented a restrictive kind of symbolic sensors working on monodimensional discrete sets of measurement, ordered lexical sets and static phenomena. We actually work on sensors using multidimensional measurement sets and non ordered lexical sets. Furthermore, an extension to a dynamic interpretation and to fuzzy symbolic informations is undergoing.

# References

[1] N.C. Burd, A.P. Dorey, "Intelligent transducers", Journal of Microcomputer Applications, No 7, pp. 87-97, 1984.
[2] J. M. Giachino, "Smart sensors", Sensors and actuators, Vol. 10, pp. 239-248, 1986.
[3] E. Bois, "Optical fibre displacement: how fibres can benefit smart sensors", Proc. of IFAC Advanced Information Processing in Automatic Control, Nancy, France, vol. 1, pp. 239-245, July 1989.
[4] J. M. Favenec, "Smart sensors in industry", Journal of physics. e. scientific instruments, No 20, pp. 1087-1090, 1987.
[5] A. Missier, N. Piera, L. Travé-Massuyès, "Order of magnitude qualitative algebras : A survey", Revue d'intelligence artificielle, Vol. 3, No 4, pp. 95-109, 1989.
[6] D. Luzeaux, "Normalizing symbolic reasoning", accepted for IMACS DSS&QR workshop, Toulouse, France, march 1991.
[7] H. Emptoz, M. Lamure, "A systemic approach to pattern recognition", Robotica, 1987, Vol. 5, p. 129-133.